\documentclass[a4paper]{jpconf}
\usepackage{graphicx}
\begin{document}
\title{Area-dependence of spin-triplet supercurrent in ferromagnetic Josephson junctions}

\author{Yixing Wang, W P Pratt, Jr and Norman O Birge}

\address{Department of Physics and
Astronomy, Michigan State University, East Lansing, Michigan
48824-2320, USA}

\ead{birge@pa.msu.edu}

\begin{abstract}
In 2010, several experimental groups obtained compelling evidence
for spin-triplet supercurrent in Josephson junctions containing
strong ferromagnetic materials. Our own best results were obtained
from large-area junctions containing a thick central Co/Ru/Co
``synthetic antiferromagnet" and two thin outer layers made of Ni
or PdNi alloy.  Because the ferromagnetic layers in our samples
are multi-domain, one would expect the sign of the local
current-phase relation inside the junctions to vary randomly as a
function of lateral position.  Here we report measurements of the
area dependence of the critical current in several samples, where
we find some evidence for those random sign variations. When the
samples are magnetized, however, the critical current becomes
clearly proportional to the area, indicating that the
current-phase relation has the same sign across the entire area of
the junctions.
\end{abstract}

The interplay between magnetism and superconductivity has been a
fertile ground for new physics over several decades.  On the one
hand, bulk materials exhibiting both forms of order are rare,
although there is recent progress in discovering new materials
with exotic properties
\cite{Niewa:352,Shermadini:117602,Jeevan:054511}. On the other
hand, combining conventional superconducting (S) and ferromagnetic
(F) materials relaxes the constraints, leading to a wide variety
of possibilities. The interplay between the two is governed either
by the magnetic field \cite{LyuksyutovReview} or the exchange
field \cite{BuzdinReview}; in this work we restrict our attention
to the latter. The essential physics arises from a modification of
Andreev reflection -- the microscopic process by which electrons
enter or leave S.  A conceptually simple picture is to consider
two electrons belonging to a Cooper pair, which happen to cross
the S/F interface. Because they enter opposite spin bands in F,
they have different Fermi wavevectors, hence the pair acquires a
center-of-mass momentum perpendicular to the S/F interface
\cite{Demler}. That, in turn, leads to oscillations in the pair
correlation function, which manifest themselves as oscillations in
the $T_c$ of S/F bilayers \cite{Jiang:95}, inversion in the
proximity-induced density of states on the F side of S/F bilayers
\cite{Kontos:01}, and oscillations between ``0" and ``$\pi$"
states in S/F/S Josephson junctions \cite{Ryazanov:01,Kontos:02}.
The oscillatory phenomena associated with S/F hybrids are
fascinating and have potential for some applications; on the other
hand, they are very short-range \cite{BuzdinReview}, decaying on a
length scale governed by the exchange energy or by the mean free
path in F.

A new chapter in S/F physics was opened in 2001 with the
prediction that spin-triplet pair correlations can be induced in
S/F hybrids made from conventional spin-singlet superconductors,
in the presence of specific types of magnetic inhomogeneity that
include non-collinear magnetizations
\cite{Bergeret:01a,Kadigrobov:01,Eschrig:03,Eschrig:2011}. Since
the two electrons from a spin-triplet pair with $m=\pm1$ (where
$m$ is the component of spin angular momentum along the
magnetization direction of F) live in the same spin band in F,
they are not subject to the exchange field. As a result, such
spin-triplet pair correlations are long-range in F, decaying on a
length scale governed by the temperature or by spin-flip or
spin-orbit scattering.  The experimental search for spin-triplet
pairs produced a few promising hints a few years ago
\cite{Keizer:06,Sosnin:06}, but conclusive evidence for their
existence came only last year when several groups reported
long-range supercurrents in S/F/S Josephson junctions
\cite{Khaire:10,Robinson:10,Sprungmann:10,Anwar:10}. Our own
results \cite{Khaire:10} were based on junctions of the form
S/N/F'/N/SAF/N/F''/N/S (see Fig. 1), where SAF is a Co/Ru/Co
``synthetic antiferromagnet" \cite{Parkin:90}, N is Cu, and F' and
F'' are thin ferromagnetic layers (Ni in this work).  Conversion
of spin-singlet to spin-triplet pairs in our samples can be viewed
as taking place in two steps \cite{Eschrig:2011}: 1) The two
electrons of a spin-singlet Cooper pair leaving the bottom S layer
acquire different phase factors in the F' layer, which generates
the $m=0$ triplet component; 2) Rotation of the $m=0$ triplet
component into the basis defined by the SAF magnetization
generates the $m=1$ and $m=-1$ triplet components in the new
basis, as long as it is non-collinear with the F' basis.
Conversion back to spin-singlets occurs between the SAF and F"
layers as the inverse process. This type of junction, with three
ferromagnetic layers, had been previously proposed theoretically
by Houzet and Buzdin \cite{Houzet:07}, but with a simple F layer
in place of the SAF, and without the crucial inner two N layers
which serve to magnetically decouple the F' and F'' layers from
the Co layers. The critical supercurrent in these samples hardly
decays as the total Co thickness varies from 12 to 30 nm
\cite{Khaire:10}, in sharp contrast to similar samples without F'
layers where the supercurrent decays very rapidly with increasing
Co thickness \cite{Khasawneh:09}.

\begin{figure}[tbh]

\includegraphics[width=3.0in]{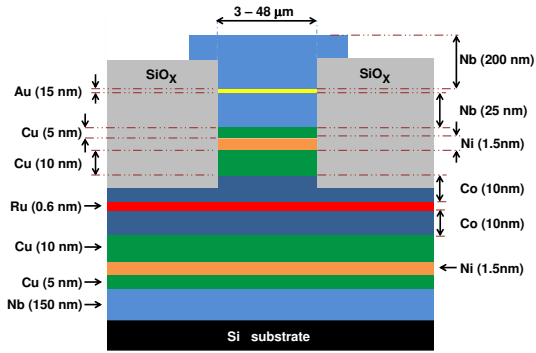}
\begin{minipage}[b]{3in}
\caption{Schematic diagram of the Josephson junctions used in the
this work, shown in cross-section. The purpose of the Au layer is
to prevent oxidation during processing; at low temperature it
becomes fully superconducting by the proximity
effect.}\label{Schematic}
\end{minipage}
\end{figure}

The results described above \cite{Khaire:10} immediately raise
several questions, only some of which have been addressed by
subsequent work \cite{Khasawneh:11}.  One of the most important
outstanding questions concerns the current-phase relation of the
junctions, or more precisely, the current-phase relation as a
function of position across the lateral dimensions of the
junction. At issue is the theoretical prediction that S/F'/F/F''/S
or S/F'/SAF/F''/S junctions (we omit explicit reference to the N
layers for simplicity) can be either ``0" or ``$\pi$" junctions
depending on the relative orientations of the magnetizations in
the three ferromagnetic layers \cite{Houzet:07, Bergeret:03,
VolkovEfetov:10, Trifunovic:10}.  (In the weak-coupling limit, a
0-junction has the current-phase relation $I_s = I_c$ sin$(\phi)$,
whereas a $\pi$-junction has the current-phase relation $I_s =
I_c$ sin$(\phi+\pi)=-I_c$ sin$(\phi)$, where $\phi$ is the
gauge-invariant superconducting phase difference between the two S
electrodes.) Looking through the junction in the direction of
current flow, the three magnetizations can either rotate in the
same direction from one to the next (with either chirality), or
the rotation from the F' layer to F can be in the opposite sense
with respect to the rotation from F to the F'' layer. (We define
the direction of rotation using the angle that is less than
$\pi$.)  In the former case, the junction is a 0-junction, whereas
in the latter case it is a $\pi$-junction \cite{Trifunovic:10}.
Since our samples have relatively large lateral dimensions (10,
20, or 40 $\mu$m in our published work \cite{Khaire:10}), and
since a typical Co domain size is only on the order of 3$\mu$m in
our sputtered films in the relevant thickness range
\cite{Borchers}, we expect our samples to contain many domains
which produce a random pattern of 0 and $\pi$ junctions.  In that
case one would expect the typical critical current of a sample to
be proportional to the square root of the number of domains, hence
proportional to the square root of the junction area. This is
quite different from the usual case where the current-phase
relation is uniform across the entire area of the junction, and
the critical current is proportional to the junction area. The
main goal of this paper is to determine experimentally which type
of area scaling occurs in our samples.

Our sample fabrication procedure is described in detail in our
previous publications \cite{Khaire:09,Khasawneh:09,Khaire:10}. All
of the samples discussed in the present work contain two
1.5-nm-thick Ni layers as the F' and F'' layers as shown in Figure
1. Photolithography and ion milling are used to define Josephson
junctions with circular cross-section. Most of our published data
were obtained from junctions with diameters of 10 or 20 $\mu$m.
Taking into account sample-to-sample fluctuations, we were not
able to determine the area scaling law from our previous data sets
\cite{Khaire:10, Khasawneh:11}. Since the goal of the present work
is to measure how the critical current scales with the junction
area, we fabricated junctions with diameters of 3, 6, 12, and 24
$\mu$m on each substrate. Although the success rate of the largest
junctions is low, we obtained data from 3, 6, and 12 $\mu$m
diameter junctions on several different substrates. All of the
measured junctions display $I-V$ characteristics obeying the
standard overdamped form: $V = R_N (I^2-I_c^2)^{1/2}$, where $R_N$
is the normal state resistance.

\begin{figure}[tbh!]

\includegraphics[width=4.0in]{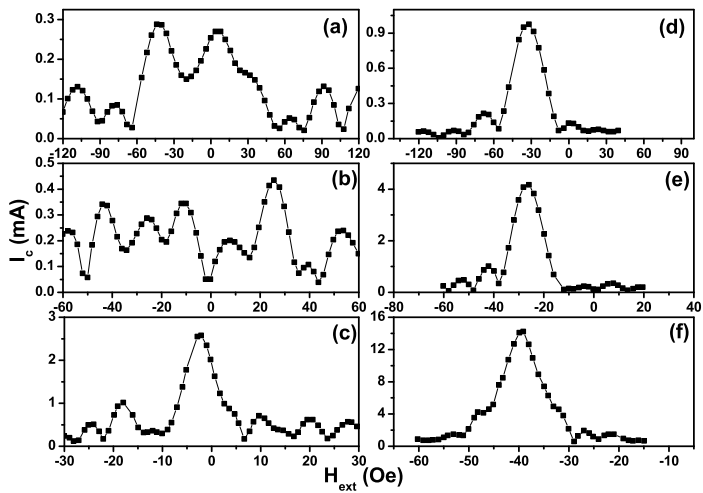}
\begin{minipage}[b]{2in}
\caption{Josephson junction critical current vs. magnetic field
applied in the plane of the substrate -- perpendicular to the
current direction. In the virgin state (panels a-c), the resulting
Fraunhofer patterns vary in quality.  After the same samples are
magnetized (panels d-f), the quality of the Fraunhofer patterns
improves and the maximum critical currents are much larger. The
data shown were taken at $T=4.2$K from junctions fabricated on a
single substrate, with junction diameters of 3, 6, and 12 $\mu$m
for the upper, middle, and lower panels, respectively.}
\label{Fraunhofers}
\end{minipage}
\end{figure}

For every Josephson junction we fabricate, we measure the
dependence of $I_c$ with magnetic field applied perpendicular to
the current direction, i.e. in the plane of the substrate. The
left panels of Figure 2 show the resulting ``Fraunhofer patterns"
of a typical set of junctions in the virgin state, fabricated on a
single substrate, with diameters of 3, 6, and 12 $\mu$m.  The
quality of the virgin-state Fraunhofer patterns varies, which is
not surprising given the amount of magnetic material contained in
the junctions \cite{Bourgeois:01, Khaire:09}.  If the SAF is
working effectively, then the Co layers produce zero net magnetic
flux in the junctions, because a given domain in one Co layer is
antiparallel to that in the other Co layer \cite{Khasawneh:09}.
Deviations from ideal Fraunhofer patterns are most probably due to
the two Ni layers. Figure 2 also shows Fraunhofer patterns of the
same three junctions after they have been subjected to a large
in-plane magnetic field of 1800 Oe. (The large field is then
removed to obtain the low-field Fraunhofer pattern.) One
immediately notices three features in the data: 1) the quality of
the Fraunhofer patterns is improved; 2) the central peak of the
pattern is shifted to a small negative field; and 3) the maximum
critical current determined from the largest peak in the pattern
is much larger than in the virgin state data. All three features
can be explained by two hypotheses: i) the Ni layers become fully
magnetized in the direction of the applied field; and ii) the SAF
undergoes a ``spin-flop" transition, whereby the two Co layers end
up perpendicular to the direction of the applied field
\cite{Zhu:98, Tong:00}. The net result is that the Ni and Co
magnetizations end up perpendicular to each other, which optimizes
generation of the spin-triplet pairs \cite{Houzet:07,
VolkovEfetov:10, Trifunovic:10}. Meanwhile, the central peak of
the Fraunhofer pattern is shifted to a small negative field where
the magnetic flux due to the external field exactly cancels the
magnetic flux due to the Ni magnetization \cite{Ryazanov:99,
Khaire:09}. Further evidence for the occurrence of the spin-flop
transition in our samples will be provided elsewhere
\cite{Klose:11}.

\begin{figure}[tbh]

\includegraphics[width=3.5in]{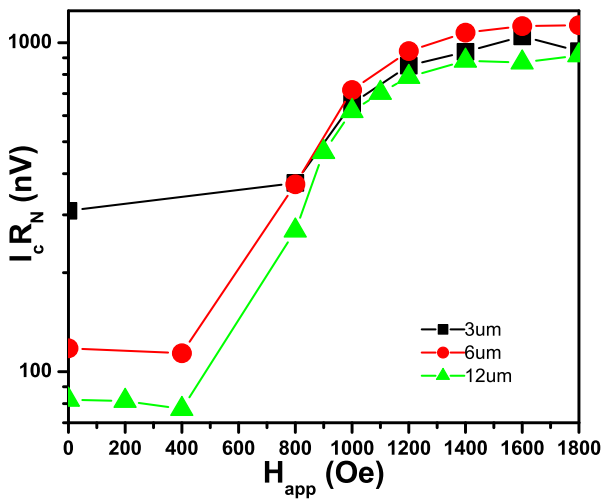}
\begin{minipage}[b]{2.5in}
\caption{Critical current times normal-state resistance ($I_cR_N$)
as a function of magnetizing field $H_{app}$ for the three
different diameter Josephson junctions shown in Figure 2.  The
large increase in $I_cR_N$ occurs in the vicinity of the coercive
field of the outer Ni layers, which is approximately 800 Oe for
the Ni thickness of $1.5$nm in these samples.  All data were taken
at $T=4.2$K.}\label{Ic-vs-H}
\end{minipage}
\end{figure}

The evolution of the $I_c$ enhancement as the samples are
magnetized is shown in Figure 3.  The figure shows $I_cR_N$ as a
function of $H_{app}$, the magnitude of the applied field, as that
field is increased in discrete steps. After each step, the full
Fraunhofer pattern is measured near zero field; the quantity
plotted is from the value of $I_c$ obtained from the peak of the
resulting Fraunhofer pattern. For all three samples, $I_c$ starts
to increase when $H_{app}$ exceeds about 800 Oe, and attains a
plateau when $H_{app}$ reaches about 1800 Oe. Data from samples
with different Ni thicknesses \cite{Klose:11} show that the values
of $H_{app}$ where $I_c$ increases vary with Ni thickness in a
manner consistent with the coercive field of thin Ni films (larger
coercive field for thinner films).  The field at which the
spin-flop transition occurs cannot be determined from these data;
all we know is that it is less than 500 Oe -- the coercive field
of our thickest Ni layers measured.

Returning to the question of how $I_c$ scales with junction area,
Figure 4 shows a plot summarizing data from 18 Josephson junctions
on 6 separate substrates.  We plot $I_cR_N$ in both the virgin
state and after the samples are fully magnetized, as a function of
junction diameter.  Several systematic features of the data stand
out.  First, the data obtained from the magnetized samples are
remarkably consistent over all junction sizes and all substrates.
Since $R_N$ is inversely proportional to the junction area $A$,
the fact that $I_cR_N$ is independent of area implies that $I_c$
scales linearly with junction area in the magnetized state, i.e.
$I_c \propto A$. This means that the supercurrent is adding
coherently across the area of the junction.  That is expected when
the two Ni layers are magnetized in the same direction across the
whole junction area.  For example, if we define the direction of
$H_{app}$ as $\theta = 0$, so that the two Ni magnetizations point
along $\theta = 0$, then the lower and upper Co
layers should point either along $\theta = \pi/2$ and
$-\pi/2$, respectively, or the converse, due to the spin-flop. In
either case, the sense of rotation from the bottom Ni layer to the
lower Co layer will be the same as the sense of rotation from the
upper Co layer to the top Ni layer.  This would be true even if
the Co magnetizations chose an angle different from $\pm \pi/2$,
as long as that angle is not $0$ and the two Co magnetizations
point in opposite directions.

The virgin state data shown in Figure 4 show more variation from
substrate to substrate.  Before discussing those data, we first
mention some variations in the ion milling process we used to
fabricate the samples. To produce Josephson junctions with a
well-defined lateral area, it is only necessary to ion mill
through the upper 25-nm-thick Nb layer (see Fig. 1). We generally
continue milling through the two Cu layers, the F' layer, and
part-way through the upper Co layer to ensure that there is no Nb
left outside of the junction area. Of the samples reported in this
study, samples \#9, \#12 and \#14 were ion milled part-way through
the upper Co layer. In these samples the lower Co layer is
unaltered by the ion milling, and should maintain the domain
structure of the continuous film. For these samples, panel (a) of
Figure 4 shows that the virgin-state $I_cR_N$ decreases with
increasing junction diameter $D$, with an almost inverse linear
dependence (except for the 12$\mu$m pillar of sample {\#12}). That
means that $I_cR_N$ varies nearly inversely with the square root
of the junction area, $I_cR_N \propto A^{-1/2}$, which means that
$I_c$ is proportional to the square root of the area, $I_c \propto
A^{1/2}$. That is what one would expect if the junctions consist
of multiple domains contributing a random distribution of 0 and
$\pi$ junctions. In this scenario, the large increase in $I_c$
that occurs when the samples are magnetized is largely due to a
transition from $I_c \propto A^{1/2}$ scaling to $I_c \propto A$
scaling. Optimization of the angle between the Ni and Co
magnetizations to $\pi/2$ also plays a factor \cite{Klose:11}; we
cannot determine from our data which factor plays the larger role.

\begin{figure}[h]
\begin{minipage}{2.8in}
\includegraphics[width=2.8in]{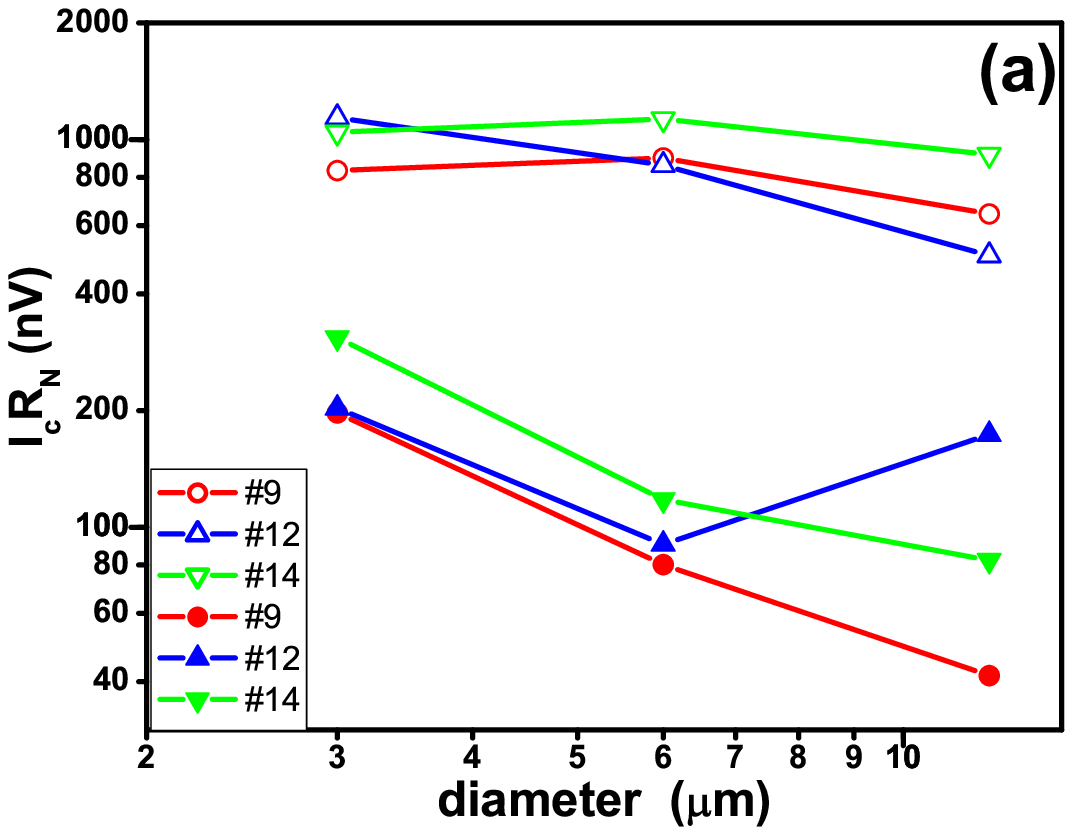}
\end{minipage}\hspace{1pc}%
\begin{minipage}{2.8in}
\includegraphics[width=2.8in]{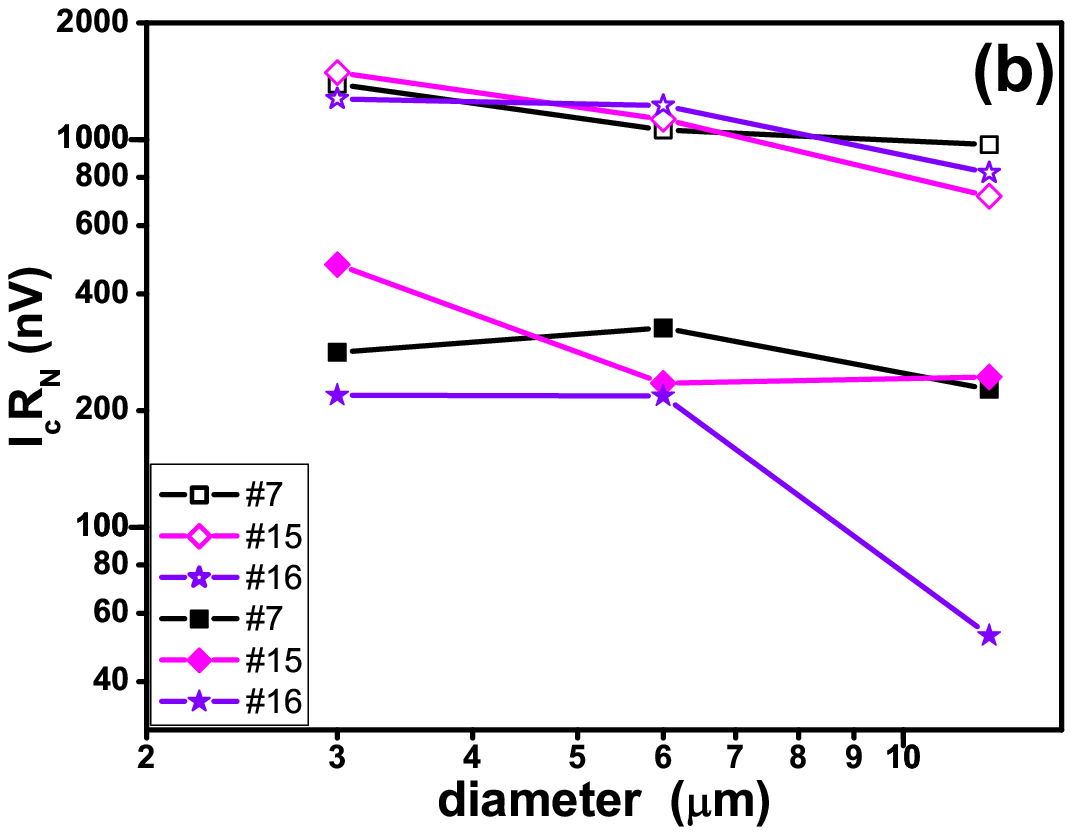}
\end{minipage}

\caption{Summary of $I_cR_N$ vs. Josephson junction lateral
diameter for all samples measured, in both the virgin state (solid
symbols) and after magnetization (open symbols). Each different
symbol shape and color refer to a different
substrate.}\label{Size-dependence}
\end{figure}

Panel (b) in Figure 4 shows data from several substrates that were
ion milled further -- either part-way through the lower Co layer
(samples \#15 and \#16) or in one case all the way through the
lower Co layer (sample \#7). In those samples the scaling of the
virgin state data is less clear, and may even be consistent with
$I_c \propto A$ scaling.  It is possible that further ion milling
of the central Co/Ru/Co SAF and even the bottom Ni layer (in the
case of sample \#7) imposes constraints on the domain structure of
the ferromagnetic layers, so that the domain magnetizations can no
longer be described as pointing in random directions.

The large sample-to-sample fluctuations in $I_cR_N$ seen in the
virgin state data in Figure 4 are strongly correlated with the
quality of the Fraunhofer patterns.  For example, the 12 $\mu$m
junction of sample \#12, which clearly deviates from the $I_c
\propto A^{1/2}$ scaling proposed above, had a particularly nice
Fraunhofer pattern -- much nicer than those from the 3 and 6
$\mu$m junctions on the same substrate.  Similar correlations were
observed in other samples: e.g. the 6 $\mu$m junction of sample
\#16 had a nicer Fraunhofer pattern than did the 3 or 12 $\mu$m
junctions on that substrate.

In conclusion, we have observed strong $I_c \propto A$ scaling in
S/F'/SAF/F''/S Josephson junctions after they are magnetized,
demonstrating coherent superposition of supercurrent over the
whole junction area. In the virgin state, there is evidence for
$I_c \propto A^{1/2}$ scaling in some samples, but not in others.
Large variations in the quality of the Frauhofer patterns acquired
from the virgin-state samples partially mask the area scaling.  We
plan to repeat this study using Pd$_{1-x}$Ni$_x$ alloy in place of
Ni for the F' and F'' layers, since our previous work showed that
PdNi layers do not distort the virgin-state Fraunhofer patterns
\cite{Khaire:09}.

Acknowledgments:  We acknowledge helpful conversations with S.
Bergeret and M. Eschrig. We also thank R. Loloee and B. Bi for
technical assistance, and use of the W.M. Keck Microfabrication
Facility. This work was supported by the U.S. Department of Energy
under Grant No. DE-FG02-06ER46341.


\section*{References}
\begin{thebibliography}{9}

\bibitem{Niewa:352} Niewa R, Shlyk L, Blaschkowski B 2011 {\it Z. Kristallogr.} {\bf 226} 352

\bibitem{Shermadini:117602} Shermadini Z \textit{et al.} 2011 {\it Phy. Rev. Lett.} {\bf 106} 117602

\bibitem{Jeevan:054511} Jeevan H S, Kasinathan D, Rosner H, Gegenwart P 2011 {\it Phys. Rev. B} {\bf 83} 054511

\bibitem{LyuksyutovReview} Lyuksyutov I F and Pokrovsky V L 2005 {\it Adv. Phys.} {\bf 54} 67

\bibitem{BuzdinReview} Buzdin A I 2005 {\it Rev. Mod. Phys.} {\bf 77} 935

\bibitem{Demler} Demler E A, Arnold G B and Beasley M R 1997 {\it Phys. Rev. B} {\bf 55} 15174

\bibitem{Jiang:95} Jiang J S, Davidovic D, Reich D H and Chien C L 1995 {\it Phys. Rev. Lett.} \textbf{74} 314

\bibitem{Kontos:01} Kontos T, Aprili M, Lesueur J and Grison X 2001 {\it Phys. Rev. Lett.} \textbf{86} 304

\bibitem{Ryazanov:01} Ryazanov V V, Oboznov V A, Rusanov A Yu, Veretennikov A V, Golubov A A and
Aarts J 2001 {\it Phys. Rev. Lett.} \textbf{86} 2427

\bibitem{Kontos:02} Kontos T, Aprili M, Lesueur J, Genet F, Stephanidis B, Boursier R 2002 {\it Phy. Rev. Lett.} {\bf 89} 137007

\bibitem{Bergeret:01a} Bergeret F R, Volkov A F and Efetov E B 2001 {\it Phys. Rev. Lett.} {\bf 86} 4096

\bibitem{Kadigrobov:01} Kadigrobov A, Shekhter R I and Jonson M 2001 {\it Europhys. Lett.} \textbf{54} 394

\bibitem{Eschrig:03} Eschrig M, Kopul J, Cuevas J C and Gerd Schön 2003 {\it Phys. Rev. Lett.} \textbf{90} 137003

\bibitem{Eschrig:2011} Eschrig M 2011 {\it Physics Today} \textbf{64}, No. 1, 43

\bibitem{Keizer:06} Keizer R S,Goennenwein S T B, Klapwijk T M, Miao G, Xiao G and Gupta A 2006 {\it Nature (London)}
\textbf{439} 825

\bibitem{Sosnin:06} Sosnin I, Cho H, Petrashov V T and Volkov A F 2006 {\it Phys. Rev. Lett.} \textbf{96} 157002

\bibitem{Khaire:10} Khaire T S, Khasawneh M A, Pratt Jr W P and Birge N O 2010 {\it Phys. Rev. Lett.} \textbf{104}
137002

\bibitem{Robinson:10} Robinson J W A, Witt J D S and Blamire M G 2010 {\it Science} \textbf{329} 59

\bibitem{Sprungmann:10} Sprungmann D, Westerholt K, Zabel H, Weides M and Kohlstedt H 2010 {\it Phys. Rev. B} \textbf{82} 060505

\bibitem{Anwar:10} Anwar M S, Czeschka F, Hesselberth M, Porcu M and Aarts J 2010 {\it Phys. Rev. B} \textbf{82} 100501

\bibitem{Parkin:90} Parkin S S P, More N and Roche K P 1990 {\it Phys.Rev. Lett.} \textbf{64} 2304

\bibitem{Houzet:07} Houzet M and Buzdin A I 2007 {\it Phys. Rev. B} \textbf{76} 060504(R)

\bibitem{Khasawneh:09} Khasawneh M A, Pratt Jr W P and Birge N O 2009 {\it Phys. Rev. B} \textbf{80} 020506(R)

\bibitem{Khasawneh:11} Khasawneh M A, Khaire T S, Klose C, Pratt Jr W P and Birge N O 2011 {\it Supercond. Sci. Technol.} \textbf{24} 024005

\bibitem{Bergeret:03} Volkov A F, Bergeret F S and Efetov K B 2003 {\it Phys. Rev. Lett.} {\bf 90} 117006

\bibitem{VolkovEfetov:10} Volkov A F and Efetov K B 2010 {\it Phys. Rev. B} \textbf{81} 144522

\bibitem{Trifunovic:10} Trifunovic L and Radovic Z 2010 {\it Phys. Rev. B} \textbf{82} 020505(R)

\bibitem{Borchers} Borchers J A, et al 1999 {\it Phys. Rev. Lett.} \textbf{82} 2796

\bibitem{Khaire:09} Khaire T S, Pratt Jr W P and Birge N O 2009 {\it Phys. Rev. B} \textbf{79} 094523

\bibitem{Bourgeois:01} Bourgeois O, Gandit P, Lesueur J, Sulpice A, Grison X and Chaussy J 2001 {\it Eur. Phys. J. B} \textbf{21} 75

\bibitem{Zhu:98} Zhu J G and Zheng Y 1998 {\it IEEE Trans. Mag.} \textbf{34} 1063

\bibitem{Tong:00} Tong H C, Qian C, Miloslavsky L, Funada S, Shi X, Liu F and Dey S2000 {\it J. Appl. Phys.} \textbf{87}
5055

\bibitem{Ryazanov:99} Ryazanov V V 1999 {\it Physics-Uspekhi} \textbf{42} 825

\bibitem{Klose:11} Klose C, Khaire T S, Wang Y, Pratt Jr W P, Birge N O, McMorran B J, Ginley T P,
Borchers J A, Kirby B J, Maranville B B, Unguris J 2011 {\it
Preprint} arXiv:1108.5666

\end{thebibliography}
\end{document}